\pdfoutput=1

\documentclass[11pt,twoside,a4paper,cmspaper,final,collab]{cms-tdr}
\def\svnVersion{124241:124242}\def\cmsCernNo{2012-042}\def\cmsCernDate{\today}\def\cmsMessage{Submitted to Physics Letters B}

\begin{document}\cmsNoteHeader{HIN-11-013}

\hyphenation{had-ron-i-za-tion}
\hyphenation{cal-or-i-me-ter}
\hyphenation{de-vices}
\RCS$Revision: 124242 $
\RCS$HeadURL: svn+ssh://alverson@svn.cern.ch/reps/tdr2/papers/HIN-11-013/trunk/HIN-11-013.tex $
\RCS$Id: HIN-11-013.tex 124242 2012-05-25 02:25:46Z alverson $

\providecommand{\rd}{\mathrm{d}}
\newcommand {\lum}    {\ensuremath{150\,\mu\mathrm{b}^{-1}}\xspace}
\newcommand {\roots}    {\ensuremath{\sqrt{s}}}
\newcommand {\rootsNN}  {\ensuremath{\sqrt{s_{_{NN}}}}}
\newcommand {\dndy}     {\ensuremath{{\rd}N/{\rd}y}}
\newcommand {\dnchdy}   {\ensuremath{{\rd}N_{\text{ch}}/{\rd}y}}
\newcommand {\dndeta}   {\ensuremath{{\rd}N/{\rd}\eta}}
\newcommand {\dnchdeta} {\ensuremath{{\rd}N_{\text{ch}}/{\rd}\eta}}
\newcommand {\dndpt}    {\ensuremath{{\rd}N/{\rd}\pt}}
\newcommand {\dnchdpt}  {\ensuremath{{\rd}N_{\text{ch}}/{\rd}\pt}}
\newcommand {\deta}     {\ensuremath{\Delta\eta}}
\newcommand {\dphi}     {\ensuremath{\Delta\phi_{1,2}}}
\newcommand {\AJ}       {\ensuremath{A_J}}
\newcommand {\npart}    {\ensuremath{N_\text{part}}}
\newcommand {\ncoll}    {\ensuremath{N_\text{coll}}}

\newcommand {\pp}    {\Pp\Pp\xspace}
\newcommand {\PbPb}  {\mbox{PbPb}\xspace}
\newcommand {\ak} {anti-$k_\text{T}$}
\newcommand {\ptlead} {\ensuremath{p_{\text{T},1}}}
\newcommand {\ptsub} {\ensuremath{p_{\text{T},2}}}
\newcommand {\ptrat} {\ensuremath{\ptsub/\ptlead}}

\newcommand{\m}{\ensuremath{\,\text{m}}\xspace}
\providecommand{\mb}{\mbox{\,\mathrm{mb}}\xspace}
\providecommand{\nb}{\mbox{\,\mathrm{mb}}\xspace}
\newcommand {\naive}    {na\"{\i}ve}
\providecommand{\PHOJET} {\textsc{phojet}\xspace}
\providecommand{\HYDJET}{\textsc{hydjet}\xspace}
\providecommand{\PYTHYD}{\PYTHIA{+}\HYDJET}

\providecommand{\PKzS}{\ensuremath{\mathrm{K^0_S}}}
\providecommand{\Pp}{\ensuremath{\mathrm{p}}}
\providecommand{\Pap}{\ensuremath{\mathrm{\overline{p}}}}
\providecommand{\PgL}{\ensuremath{\Lambda}}
\providecommand{\PagL}{\ensuremath{\overline{\Lambda}}}
\providecommand{\PgS}{\ensuremath{\Sigma}}
\providecommand{\PgSm}{\ensuremath{\Sigma^-}}
\providecommand{\PgSp}{\ensuremath{\Sigma^+}}
\providecommand{\PagSm}{\ensuremath{\overline{\Sigma}^-}}
\providecommand{\PagSp}{\ensuremath{\overline{\Sigma}^+}}

\newlength\cmsFigWidth
\ifthenelse{\boolean{cms@external}}{\setlength\cmsFigWidth{0.75\textwidth}}{\setlength\cmsFigWidth{\textwidth}}
\ifthenelse{\boolean{cms@external}}{\providecommand{\cmsLeft}{top}}{\providecommand{\cmsLeft}{left}}
\ifthenelse{\boolean{cms@external}}{\providecommand{\cmsRight}{bottom}}{\providecommand{\cmsRight}{right}}

\hyphenation{off-line}

\cmsNoteHeader{HIN-11-013} 
\title{Jet momentum dependence of jet quenching in PbPb collisions
at $\sqrt{s_{_{NN}}}=2.76$\TeV}

\date{\today}

\abstract{Dijet production in PbPb collisions at a nucleon-nucleon center-of-mass
energy of 2.76\TeV is studied with the CMS detector at the LHC.
A data sample corresponding to an integrated luminosity of $150\,\mu$b$^{-1}$ is analyzed. Jets are reconstructed using combined information from tracking and calorimetry, using the anti-$k_\text{T}$ algorithm with $R$ = 0.3.
The dijet momentum
balance and angular correlations are studied
as a function of collision centrality and leading jet transverse momentum.
For the most peripheral \mbox{PbPb} collisions, good agreement of the dijet momentum balance
distributions with \mbox{pp} data and reference calculations at the same collision energy
is found, while more central collisions
show a strong imbalance of leading and subleading jet transverse momenta attributed
to the jet-quenching effect.
The dijets in central collisions are found to be more unbalanced than the reference, for leading
jet transverse momenta up to the highest values studied.
}

\hypersetup{%
pdfauthor={CMS Collaboration},%
pdftitle={Jet momentum dependence of jet quenching in PbPb collisions at sqrt(s[NN]) = 2.76 TeV},%
pdfsubject={CMS},%
pdfkeywords={CMS, physics, heavy ion, jet}}

\maketitle 

\section{Introduction}
\label{sec:introduction}

Quantum chromodynamics (QCD) predicts that a new form of matter, consisting of
deconfined quarks and gluons, is formed at very high temperatures. Calculations
in lattice QCD~{\cite{Karsch:2003jg} indicate that the transition to this
so-called quark-gluon plasma should occur at a critical temperature of around 150--175\MeV, corresponding to an energy density of about 1~GeV/fm$^3$.
Experiments have provided evidence that dense matter at high temperature can be
created in relativistic heavy ion
collisions~\cite{Adcox:2004mh,Adams:2005dq,Back:2004je,Arsene:2004fa}.

Studies of particle production at high transverse momentum (\pt) are a well-established way to
probe the properties of the medium formed in heavy-ion collisions. The yields and correlations of high momentum particles are modified through the ``jet-quenching'' effect,
resulting from the energy loss suffered
by hard-scattered partons passing through the medium~\cite{Bjorken:1982tu}.
Fast parton energy loss provides key information on the thermodynamic and transport properties of the medium traversed~\cite{CasalderreySolana:2007zz,dEnterria:2009am}. Evidence for jet quenching was first observed in the suppression of inclusive high-$\pt$ hadron production and the modification of high-$\pt$ dihadron angular correlations in nucleus-nucleus collisions at the Relativistic Heavy Ion Collider, in comparison to proton-proton collisions ~\cite{Adams:2005dq,Adcox:2004mh,Arsene:2004fa,Back:2004je}.

Recent results at the LHC~\cite{Chatrchyan:2011sx,Aad:2010bu,Alice:dihadron,HIN-10-005,Aamodt:2010jd}, using
fully reconstructed jets, correlations between jets and single particles, and charged particle measurements,
provide detailed information on the jet-quenching effect. For central collisions,
a large broadening of the dijet momentum asymmetry distributions is observed, consistent with
theoretical calculations that involve differential energy loss of back-to-back hard-scattered
partons as they traverse the medium~\cite{He:2011pd,Young:2011qx,Qin:2010mn}.
At the same time, angular correlations between the jets are found to be almost unchanged, ruling
out single-hard-gluon radiation as the leading energy loss mechanism. Studies of jet-hadron correlations,
involving vector summation of charged hadron momenta,
find that the energy balance in events with large dijet asymmetry is recovered on average
by an excess of low-momentum particles in the hemisphere of the away-side jet,
at large angles relative to the jet axes~\cite{Chatrchyan:2011sx}. These results constrain the mechanism of parton energy loss~\cite{CasalderreySolana:2010eh,Chesler:2011nc}.
Further understanding of this mechanism requires the measurement of the \pt dependence of the
observed effects.

The dijet analysis presented in this paper uses a large dataset of PbPb collisions
at a nucleon-nucleon center-of-mass energy of $\rootsNN=2.76\TeV$, collected with the Compact Muon Solenoid
(CMS) detector in 2011.
The CMS detector has a solid-angle acceptance of nearly 4$\pi$ and is designed to measure jets and energy flow, an excellent feature for studying heavy ion collisions.
With a total integrated luminosity of 150$\pm$8 $\mu$b$^{-1}$ this
dataset provides a significantly larger data sample than the 6.8$\,\mu$b$^{-1}$ analyzed in 2010, allowing an extension of previous studies to more peripheral collision
events and to jets of transverse momenta in excess of 350\GeVc.
Additionally, this paper also presents  a dijet analysis of pp collisions recorded at $\sqrt{s}=2.76$\TeV by CMS in 2011, with a total integrated
luminosity of 231\nbinv.

\section{Experimental method}
\label{sec:experimental_method}
\subsection{The CMS detector}
\label{sec:cms}

The CMS detector is described in detail
elsewhere~\cite{bib_CMS}.  The calorimeters provide hermetic coverage over a large
range of pseudorapidity, $|\eta| < 5.2$, where $\eta = -\ln [
\tan(\theta/2)]$ and $\theta$ is the polar angle relative to the counterclockwise
ion beam (the $z$ axis).
A steel and quartz-fiber Cherenkov
calorimeter, called hadron forward (HF), covers the high pseudorapidity range
$3<|\eta|<5.2$ and is used to determine the centrality of the PbPb collision.
Hadron calorimeter (HCAL) cells are grouped in projective
towers of granularity
$\Delta \eta \times \Delta \phi =0.087\times0.087$ (where $\phi$ is the azimuthal angle) at central pseudorapidities, having a segmentation about twice as large at forward
pseudorapidities. The electromagnetic calorimeter (ECAL) has a further segmentation of 5$\times$5 within a tower, and the signals in these cells are clustered together to reconstruct photons. The central calorimeters are embedded in a 3.8\unit{T} axial magnetic field produced by a superconducting solenoid.
The CMS tracking system, located inside the calorimeter, consists of silicon-pixel and silicon-strip layers covering  $|\eta| <
2.5$. This analysis uses tracks reconstructed down to transverse momenta of $900$\MeVc in PbPb collisions, with
a track momentum resolution of about 1\% at $\pt = 100$\GeVc. At high \pt, the efficiency of the tracking is not strongly \pt\ dependent, and for 100\GeVc tracks it varies from 65\% in peripheral collisions to 60\% in central collisions. A set of scintillator tiles, the beam scintillator counters (BSC), used for triggering and beam-halo rejection, is mounted on the inner side of the HF calorimeters.
The detailed Monte Carlo (MC) simulation of the CMS detector response is based on {\GEANTfour}
\cite{bib_geant}.

\subsection{Jet reconstruction in PbPb collisions}
\label{sec:jet_reconstruction}

Jets are reconstructed using the CMS ``particle-flow" algorithm~\cite{CMS-PAS-PFT-10-002,MattPFlow}.
This algorithm attempts to identify all stable particles in an
event (electrons, muons, photons, charged and neutral hadrons)
by combining information from all sub-detector systems.
The \ak\ sequential recombination algorithm, as encoded in the
{\textsc{FastJet}} framework, is used to combine the particle-flow candidates
into jets using a distance parameter R = 0.3~\cite{Cacciari:2008gp}.

The small value of R helps to reduce the deterioration of the jet energy
resolution in PbPb collisions due to
fluctuations of the background from soft interactions.
The underlying
background from soft collisions is subtracted using the same method
as employed in~\cite{Chatrchyan:2011sx} and originally described in~\cite{Kodolova:2007hd}.
This algorithm is a variant of an iterative ``noise/pedestal subtraction''
technique, where the mean and dispersion of the energies detected in rings of constant $\eta$ are subtracted from the jet. The jet and background energies are formed from the total energy, determined by particle-flow, within projective towers with the same segmentation as the HCAL, as described in Section~\ref{sec:cms}.

The jets reconstructed with the procedure above are then corrected to final state particle jets.
CMS uses a factorized multi-step approach to correct the jet energies~\cite{Chatrchyan:2011ds}. For this analysis, jet energy
corrections are derived from \PYTHIA~\cite{bib_pythia} simulations without PbPb underlying events. Jets reconstructed
with the particle-flow algorithm using heavy-ion tracking~\cite{HIN-10-005} require different corrections
than those derived with the tracking algorithm optimized for pp data, due to the difference of tracking efficiencies.

The pp sample is reconstructed with the same tracking, particle flow and jet algorithms as those used in the analysis of the PbPb data. Although the underlying event and pile-up in pp collisions at $\sqrt{s}=2.76$\TeV is small enough not to require a subtraction method, the jet algorithm works successfully in both types of environment.

\subsection{Data samples and triggers}

For online event selection,
CMS uses a two-level trigger system:
a hardware-based level-1 trigger and a software-based high level trigger (HLT). The events used in this analysis are
selected by an inclusive single-jet trigger that required
a calorimeter based jet reconstructed in HLT with  $\pt > 80$\GeVc , where the jet $\pt$ value is
corrected for the $\pt$-dependent calorimeter energy response.
The trigger efficiency is defined as the fraction
of triggered events out of a sample of minimum bias events (described below)
in bins of offline reconstructed leading-jet $\pt$.
The trigger becomes fully efficient for collisions with a leading particle-flow jet
with corrected $\pt$ greater than 100\GeVc.

In addition to the jet data sample, a minimum bias event sample was collected using
coincidences between the trigger signals from both the $+z$ and $-z$ sides of either the BSC
or the HF, which was pre-scaled to record only about 0.1--0.2\% of the collisions delivered by the LHC.
In order to suppress non-collision-related noise, cosmic-ray muons,
out-of-time triggers, and beam backgrounds, the minimum bias and jet triggers used in this analysis
were required to arrive in time with the presence of both colliding ion bunches in the interaction region.
The events selected by the jet trigger described above also satisfy all triggers and selections
imposed for minimum bias events.

\subsection{Event selection and centrality determination}
\label{sec:event_selection}
A sample of inelastic hadronic collisions is selected offline from the triggered events. Contamination from beam-halo events is removed based upon the timing of the $+z$ and $-z$ BSC signals. A requirement of a reconstructed primary collision vertex based on at least two tracks with transverse momenta above $75$~\MeVc is imposed. This requirement removes other beam related background events (e.g., beam-gas, ultraperipheral collisions) with large HF energy deposits but very few pixel detector hits. The vertex is required to be compatible with the length of the pixel clusters reconstructed in the event, as a standard method in CMS~\cite{Khachatryan:2010us}. Finally, an offline HF coincidence is applied, which requires at least three towers on each side of the interaction point in the HF with at least 3~GeV total deposited energy per tower. This event selection, including the minimum bias trigger, has an efficiency of 97\% with an uncertainty of 3\% for hadronic inelastic PbPb collisions. This efficiency is
taken into account in the centrality determination, and the uncertainty of the efficiency has a negligible effect
on the results of this study.

Table~\ref{evselcuts} shows the number of events remaining after the various selection criteria are applied. Events with a jet trigger of $\pt > 80$\GeVc are selected, followed by the offline event selection for inelastic hadronic collisions (described above). Prior to jet finding on
the selected events, a small contamination of
noise events from the electromagnetic calorimeter and hadron calorimeter is removed using signal
timing, energy distribution, and pulse-shape information \cite{ref:EGM-10-002,Chatrchyan:2009hy}. The leading and subleading jets are determined among the jets with pseudorapidity $|\eta| < 2$, which are reconstructed as described in Section~\ref{sec:jet_reconstruction}.
Events are then selected if the corrected jet \pt is larger than  $120$\GeVc (corrected
for the $\pt$- and $\eta$-dependent detector energy response). The subleading jet in the
event is required to have a corrected jet $\pt > 30$\GeVc.
The azimuthal angle between the leading and the subleading jets is required to be at least $2\pi/3$.
Further jets found in the event, beyond the leading and the subleading ones, are not considered in this analysis.
In order to remove events with residual HCAL noise that are missed by the noise-rejection algorithms,
either the
leading or subleading jet is required to have at least one track of $\pt > 4$\GeVc. For high-\pt jet events
this selection does not introduce any significant bias on the sample and removes only 2\% of the
selected dijet events.

The centrality of the collisions is represented by the number of participating nucleons (\npart) in a collision, which is correlated with the total transverse energy measured in HF. The minimum bias event sample is divided into constant fractions of total inelastic cross section and for each fraction the average value of \npart\ is determined using a Glauber calculation~\cite{Miller:2007ri}. The dispersion of the \npart\ values due to reconstruction effects is based on {\GEANTfour} simulations of events generated with a multi-phase transport {\textsc{ampt}} simulation~\cite{Lin:2004en}.

\begin{table*}[htbp]
\begin{center}
\caption{The effects of various selections applied to the data sample. In the third column, the
fractional values are with respect to the line above and in the fourth column they are with respect to the triggered sample. The selections are applied in
sequence.}
\label{evselcuts} \begin{tabular}{|l|r|r|r|}
\hline
Selections & Events remaining & \% of previous & \% of triggered \\
\hline
Jet triggered events ($\pt^{\text{corr}}>80$\GeVc) & 369\,938 & 100.00 & 100.00  \\
Offline collision selection  & 310\,792 & 84.01 & 84.01  \\
HCAL and ECAL noise rejection   & 308\,453 & 99.25 & 83.38  \\
Leading jet $\ptlead>120$\GeVc  & 55\,911 & 18.13 & 15.11  \\
Subleading jet $\ptsub>30$\GeVc    & 52\,694 & 94.25 & 14.24  \\
\dphi $>2\pi/3$ &  49\,993 & 94.87 & 13.51  \\
Track within a jet &   49\,054 & 98.12 & 13.26  \\
\hline
\end{tabular}
\end{center}
\end{table*}

\subsection{Simulated data samples}
\label{sec:pythia_samples}

In PbPb collisions there is a high multiplicity of soft particles produced, the PbPb underlying event. It is essential to understand how the jet reconstruction is modified in PbPb collisions at different centralities. This is studied with simulations of dijet events in pp collisions with the \PYTHIA event generator (version 6.423, tune Z2)~\cite{bib_pythia},
modified for the isospin content of the colliding nuclei. A minimum hard-interaction scale ($\hat{p}_\mathrm{T}$) selection of 80\GeVc\ is used to increase the number of dijet events produced in the momentum range studied. \PYTHIA simulations at lower $\hat{p}_\mathrm{T}$ (discussed in~\cite{Cacciari:2011tm}) are also investigated and found to agree with the $\hat{p}_\mathrm{T} > 80$\GeVc\ results within the uncertainties. To model the PbPb background, minimum bias \PbPb\ events are simulated with the \HYDJET event generator~\cite{Lokhtin:2005px}, version 1.8 (denoted \PYTHYD in this paper).
The parameters of \HYDJET are tuned to reproduce the total particle multiplicities, charged hadron spectra, and elliptic flow at all centralities, and to approximate
the underlying event fluctuations seen in data, differences being within the underlying event systematic uncertainty.

The full detector simulation and analysis chain is used to process both \PYTHIA dijet events and \PYTHIA dijet events embedded into \HYDJET events. The reconstruction of particle flow jets is studied by using the \PYTHIA generator jet information in comparison to the same fully reconstructed jet in \PYTHYD, matched in momentum space. The effects of the PbPb underlying event on jet \pt\ and position resolution, jet \pt\ scale, and jet-finding efficiency are determined as a function of collision centrality and jet \pt. These effects do not require corrections on the results but contribute to the systematic uncertainties.
\section{Results}
\label{sec:results}

The goal of this analysis is to characterize possible modifications of dijet event properties as a function
of centrality and leading jet transverse momentum in \PbPb\ collisions.
The analysis is performed in six bins of collision centrality: 0--10\%, 10--20\%, 20--30\%, 30--50\%, 50--70\%, and 70--100\%, the latter being the most peripheral bin. The 0--20\% most central events are further analyzed in bins of leading jet $\pt$: 120--150, 150--180, 180--220, 220--260, 260--300, 300--500\GeVc.
Throughout the paper, the results obtained
from \PbPb\ data are compared to references based on the \PYTHYD
samples described in Section~\ref{sec:pythia_samples}. The subscripts $1$ and $2$ in the kinematical quantities always refer to the leading and subleading jets, respectively.

\subsection{Dijet azimuthal correlations}
\label{sec:dphi}

\begin{figure*}[htbp]
\begin{center}
\includegraphics[width=\cmsFigWidth]{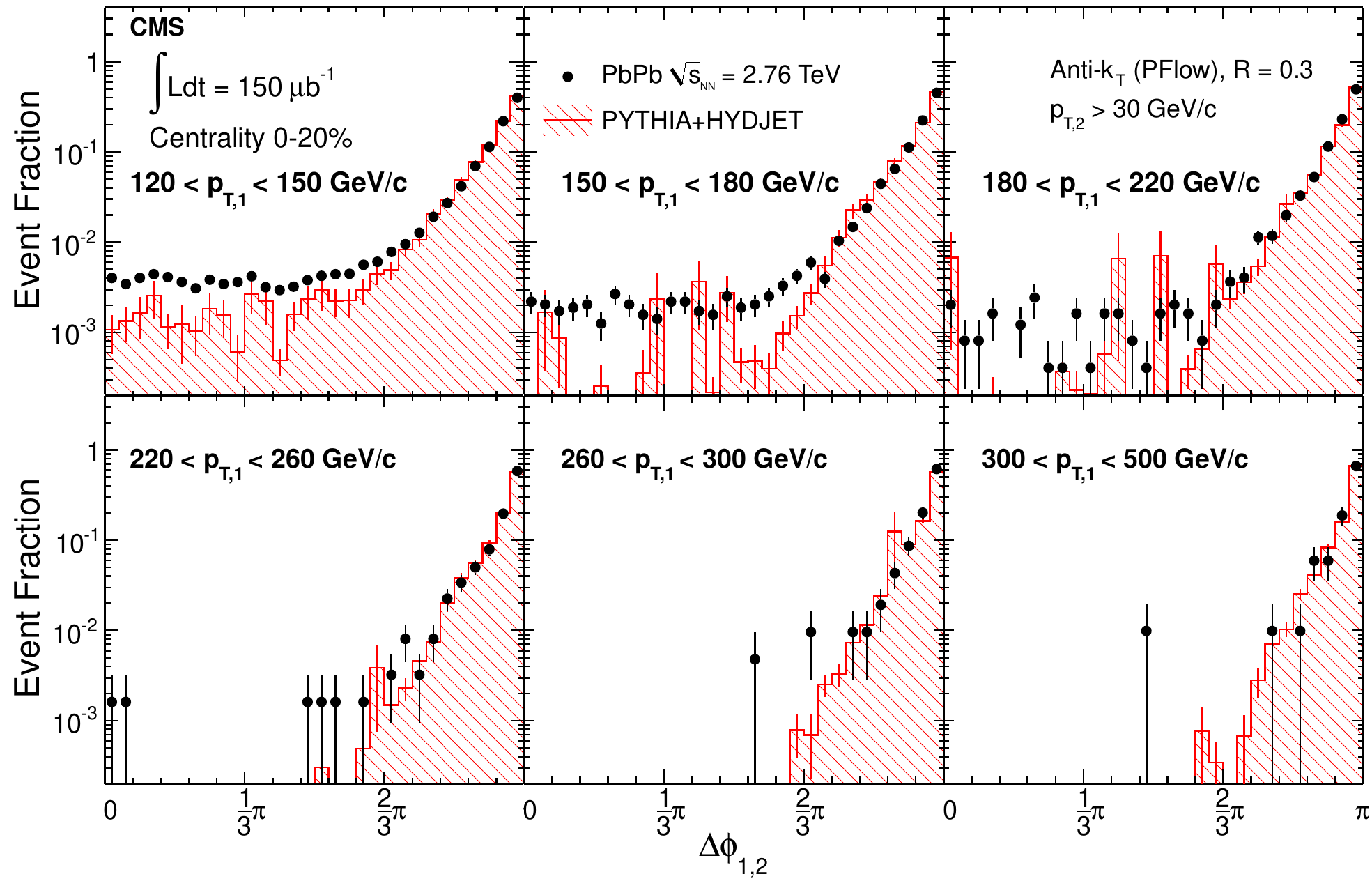}
\end{center}
\caption{Distribution of the angle $\dphi$ between the leading and subleading jets
in bins of leading
jet transverse momentum from
120 $ < \ptlead < 150$\GeVc\ to $\ptlead > 300$\GeVc\ for
  subleading jets of $\ptsub> 30$\GeVc.
Results for 0--20\% central PbPb events are shown as points while the histogram
shows the results for
{\sc {pythia}} dijets embedded into \HYDJET PbPb simulated events. The error bars represent the statistical uncertainties.}
\label{fig:dphiPt}
\end{figure*}

Earlier studies of the dijet events in heavy-ion collisions~\cite{Chatrchyan:2011sx,Aad:2010bu} have shown persistency in dijet azimuthal correlations despite the asymmetry in dijet momenta. This aspect is crucial
in the interpretation of energy loss observations~\cite{CasalderreySolana:2011rq}.
To understand the momentum dependence of the quenching effects, this study investigates the angular correlation, \ie, the opening azimuthal angle, \dphi, between the leading and subleading jets of the events, in bins of leading jet $\ptlead$.

For events with 0--20\% centrality, two features are visible in the \dphi\ distributions shown in Fig.~\ref{fig:dphiPt}: a peaking structure at $\dphi = \pi$, and a constant offset from zero in the overall distribution. The distribution around the $\dphi = \pi$ peak reflects the back-to-back dijet production and although this distribution changes across the various leading-jet \pt\ bins, there is no significant difference between PbPb data and the \PYTHYD sample.
This observation confirms the conclusions of earlier studies~\cite{Chatrchyan:2011sx,Aad:2010bu}, extending the analysis to differential leading-jet \pt bins.
The event fraction that extends to small \dphi\ values is likely due to the matching of the leading jet with a random underlying event fluctuation instead of the true subleading jet partner. The difference in the rate of such events between the PbPb data and the \PYTHYD sample is compatible with the effect of quenching, which makes it easier for a background fluctuation to supersede a genuine low \pt jet.
The fraction of these background events strongly depends on the centrality and leading jet \pt. For the purposes of the study presented in this paper, the contribution of these background events to the results is subtracted by using the events at small \dphi.

\subsection{Dijet momentum balance}
\label{sec:asymmetry}
To characterize the dijet momentum balance (or imbalance) quantitatively, we use the asymmetry ratio
\begin{equation}
\label{eq:aj}
A_J = \frac{\ptlead-\ptsub}{\ptlead+\ptsub}~.
\end{equation}
Dijets are selected with $\dphi > 2\pi/3$.
It is important to note that the subleading jet $\ptsub > 30$\GeVc\ selection imposes
a $\ptlead$-dependent limit on the magnitude of \AJ. The distributions are normalized to
the number of selected dijet events.

As discussed in Section~\ref{sec:dphi}, the contribution of background fluctuations is estimated from
the events with dijets of $\dphi < \pi/3$, and the distributions obtained from these events are subtracted from the results.
The estimated fraction of background events, as a function of both leading jet \pt\ and centrality, is shown in the bottom row of Fig.~\ref{fig:SubRate}.
The fraction of dijet events in which the subleading jet is found within the acceptance, after the subtraction of
background events, is shown in the top row of Fig.~\ref{fig:SubRate}. The events in which the subleading jet is not found should be taken into account when comparing the asymmetry distributions, although the bias is negligible for bins of leading jet $\pt > 180$\GeVc.

\begin{figure*}[htb]
\begin{center}
\includegraphics[width=\cmsFigWidth]{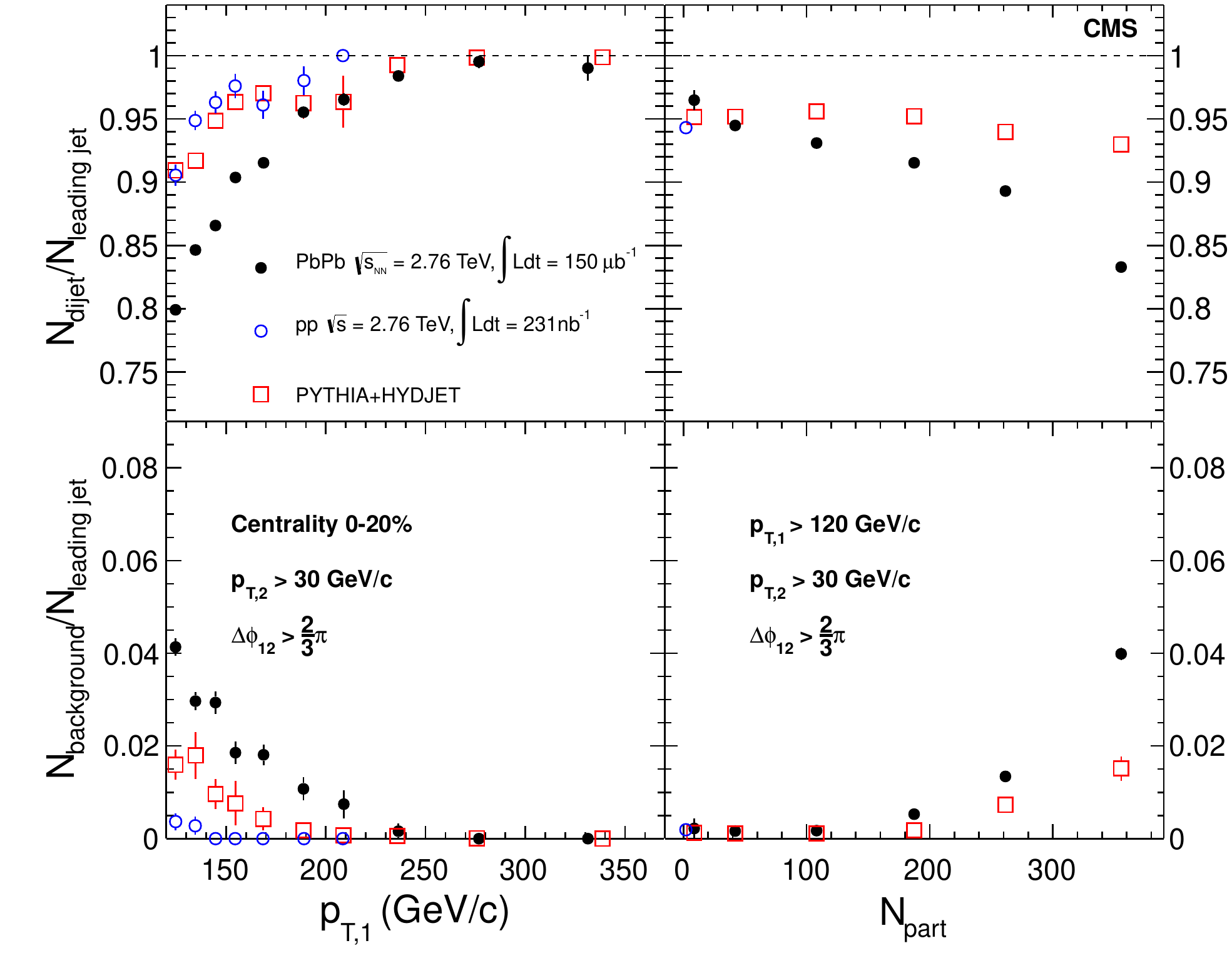}
\caption{
Fraction of events with a genuine subleading jet with $\dphi > 2\pi/3$, as a function of leading jet $\ptlead$ (left)
and \npart\ (right). The background due to underlying event
fluctuations is estimated from $\dphi < \pi/3$ events and subtracted from the number of dijets.
The fraction of the estimated background is shown in the bottom panels.
The error bars represent the statistical uncertainties.}
\label{fig:SubRate}
\end{center}
\end{figure*}

\begin{figure*}[htbp]
\begin{center}
\includegraphics[width=\cmsFigWidth]{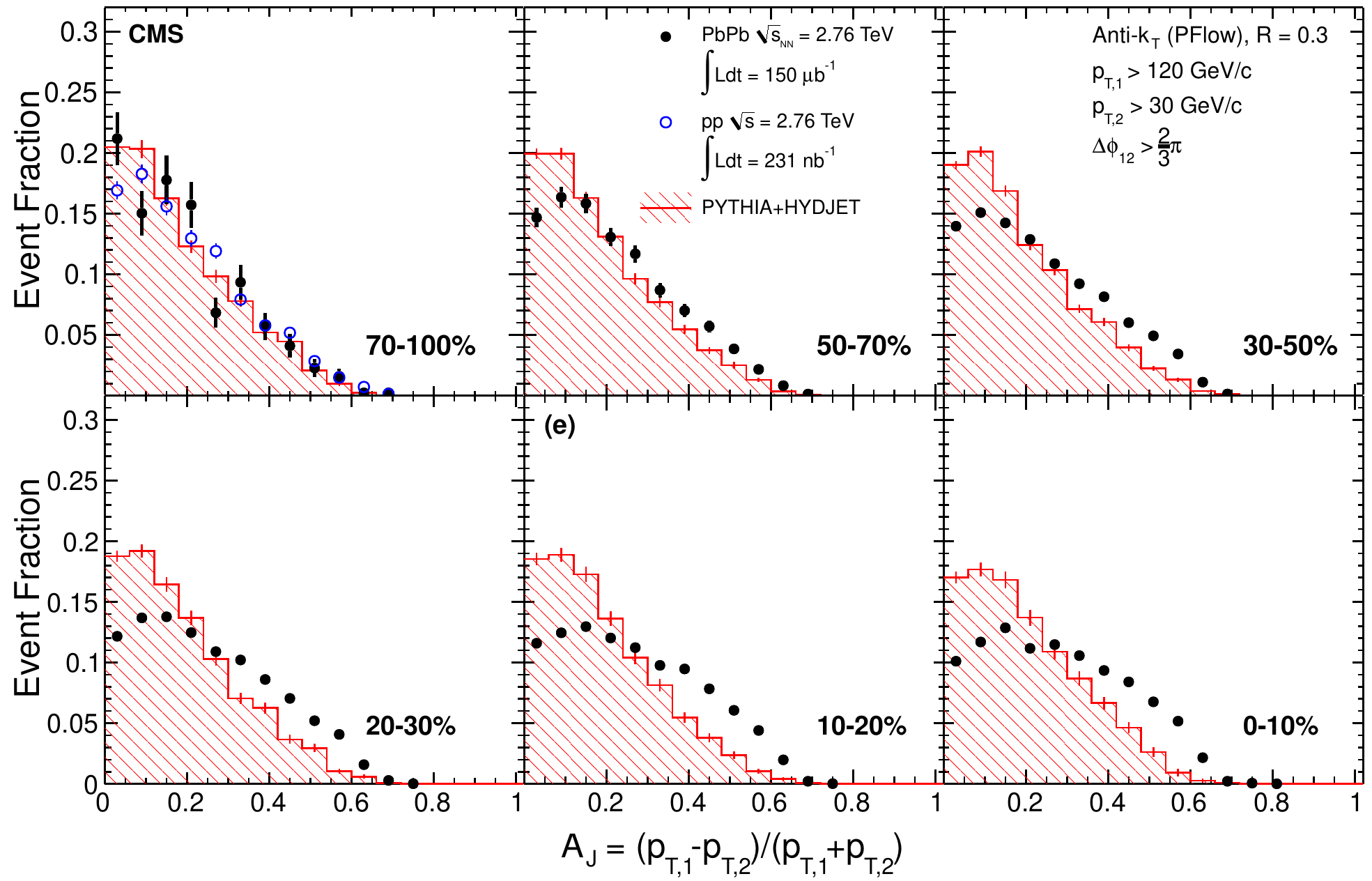}
\caption{Dijet asymmetry ratio, $A_{J}$, for leading jets of $\ptlead> 120$\GeVc\ and
  subleading jets of $\ptsub> 30$\GeVc\ with a selection of $\dphi>2\pi/3$ between the two jets.
Results are shown for six bins of collision centrality, corresponding to selections of 70--100\% to 0--10\%  of the total inelastic cross section.
Results from data are shown as points, while the histogram shows the results
for
\PYTHIA dijets embedded into \HYDJET PbPb simulated events.
Data from pp collisions at 2.76\TeV are shown as open points in comparison to PbPb results of 70--100\% centrality.
The error bars represent the statistical uncertainties.}
\label{fig:JetAsymm}
\end{center}
\end{figure*}

\begin{figure*}[htbp]
\begin{center}
\includegraphics[width=\cmsFigWidth]{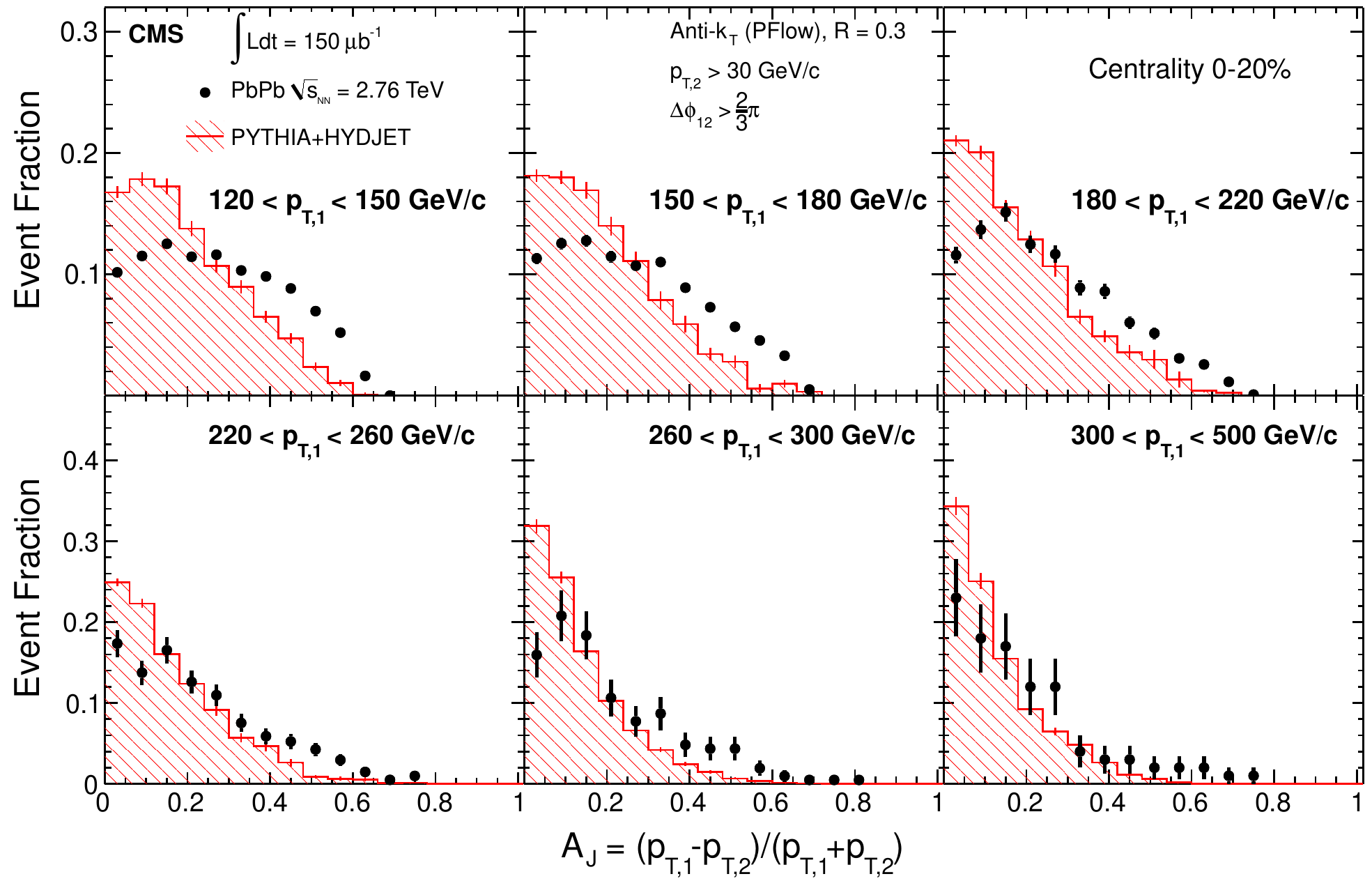}
\caption{Dijet asymmetry ratio, $A_{J}$, in bins of leading jet transverse momentum from
120 $ < \ptlead < 150$\GeVc\ to $\ptlead > 300$\GeVc\ for
  subleading jets of $\ptsub> 30$\GeVc\
and $\dphi>2\pi/3$ between leading and subleading jets.
Results for 0--20\% central PbPb events are shown as points, while the histogram
shows the results for
\PYTHIA dijets embedded into \HYDJET PbPb simulated events. The error bars represent the statistical uncertainties.}
\label{fig:JetAsymmPt}
\end{center}
\end{figure*}

\begin{figure*}[htbp]
\begin{center}
\includegraphics[width=\cmsFigWidth]{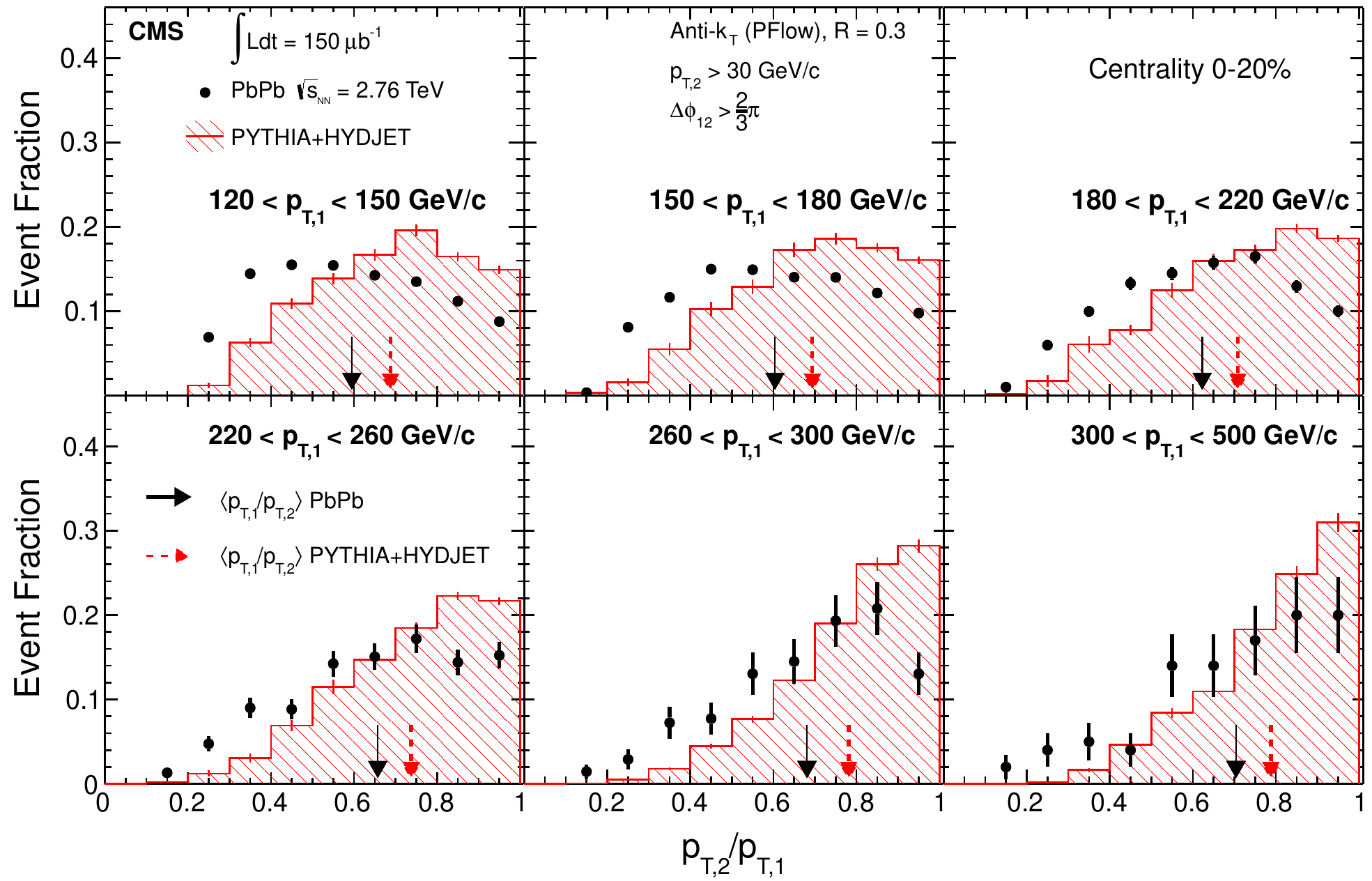}
\caption{Subleading jet transverse momentum fraction ($\ptsub/\ptlead$), in bins of leading
jet transverse momentum from
120 $ < \ptlead < 150$\GeVc\ to $\ptlead > 300$\GeVc\ for
  subleading jets of $\ptsub> 30$\GeVc\
and $\dphi>2\pi/3$ between leading and subleading jets.
Results for 0--20\% central PbPb events are shown as points, while the histogram
shows the results for
\PYTHIA dijets embedded into \HYDJET PbPb simulated events.
The arrows show the mean values of the distributions and the error bars represent the statistical uncertainties.}
\label{fig:JetFractionPt}
\end{center}
\end{figure*}

The centrality dependence of $A_J$ for \PbPb\ collisions is shown in
Fig.~\ref{fig:JetAsymm}, in comparison to results from \PYTHYD simulations.
The most peripheral events are also compared
to results from \pp\ collisions at $\sqrt{s} = 2.76\TeV$, where the same jet algorithm is used.
This comparison supports the use of the
\PYTHYD sample as a reference for the dijet asymmetry, which also takes into account underlying event
effects when comparing with PbPb data.
The shape of the dijet momentum balance distribution experiences a gradual change with collision centrality,
towards more imbalance. In contrast, the \PYTHIA simulations only
exhibit a modest broadening, even when embedded in the highest multiplicity
\PbPb\ events.

To study the momentum dependence of the amount of energy loss,
Fig.~\ref{fig:JetAsymmPt} presents the distributions of $A_J$ in different bins of leading jet \pt,
for 0--20\% central events. One observes a strong evolution in the shape of the distribution across the
various \pt\ bins, while a significant difference between PbPb data and \PYTHYD simulations persists in
each \pt\ bin. The distributions of the \ptrat\ ratio, shown in Fig.~\ref{fig:JetFractionPt}, provide a more intuitive way of quantifying the energy loss.
Both the $A_J$ and \ptrat\ distributions are affected by the cut on the subleading jet $\pt$, which should be taken into account in the interpretation of the average value. However, in the bins with leading jet $\pt> 180\GeVc$, more than 95\% of the leading jets are correlated with a subleading jet, indicating that the bias due to dijet selection is very small.

\subsection{The dependence of dijet momentum imbalance on the \texorpdfstring{\pt}{pt} of the leading jet }

The dependence of the energy loss on the leading jet momentum can be studied using the jet transverse momentum ratio
$\ptsub/\ptlead$.
The mean value of this ratio is presented as a function of $\ptlead$
in Fig.~\ref{deltaPt} for three bins of collision centrality, 50--100\%, 20--50\%, and 0--20\%.
The \PYTHYD simulations are shown as squares and the
PbPb data are shown as points. Statistical and systematic uncertainties are plotted as error bars and brackets, respectively.
The main contributions to the systematic uncertainty in $\ptsub/\ptlead$
are the uncertainties in the $\pt$-dependent residual energy scale and the effects of the underlying event on the jet energy resolution.
Earlier studies of jet-track correlations~\cite{Chatrchyan:2011sx} have shown that the energy composition of the quenched jets was not significantly different, which
puts a constraint on the energy scale uncertainty. The uncertainty on the energy scale is derived from
three sources: the uncertainty evaluated in the pp studies \cite{Chatrchyan:2011ds},
the energy scale difference in pp data and MC, and the energy scale and its parton type dependence~\cite{MattPFlow} in simulations of PbPb events (see Section~\ref{sec:pythia_samples}). These contributions are added in quadrature to assign the total uncertainty on the jet energy scale.  Using this value as a boundary, the uncertainty in the $\ptsub/\ptlead$
results is then estimated by varying the jet response at low \pt\ and at high \pt\ independently.
The uncertainty on the underlying event effects is estimated from the full
difference between \pp and \PYTHYD.
These effects add up to 6\% in the most central events.
For the low leading-jet \pt\ bins, jet reconstruction efficiency also introduces a minor uncertainty on the order of 1\%.
Uncertainties due to additional misreconstructed jets, calorimeter noise, and the track requirement are negligible compared
to the dominating sources of uncertainty.
For the centrality bins of 50--100\%, 20--50\% and 0--20\%, the sources of systematic uncertainty are summarized in Table~\ref{RatioSystematics}.

\begin{figure*}[htbp]
\begin{center}
\includegraphics[width=\cmsFigWidth]{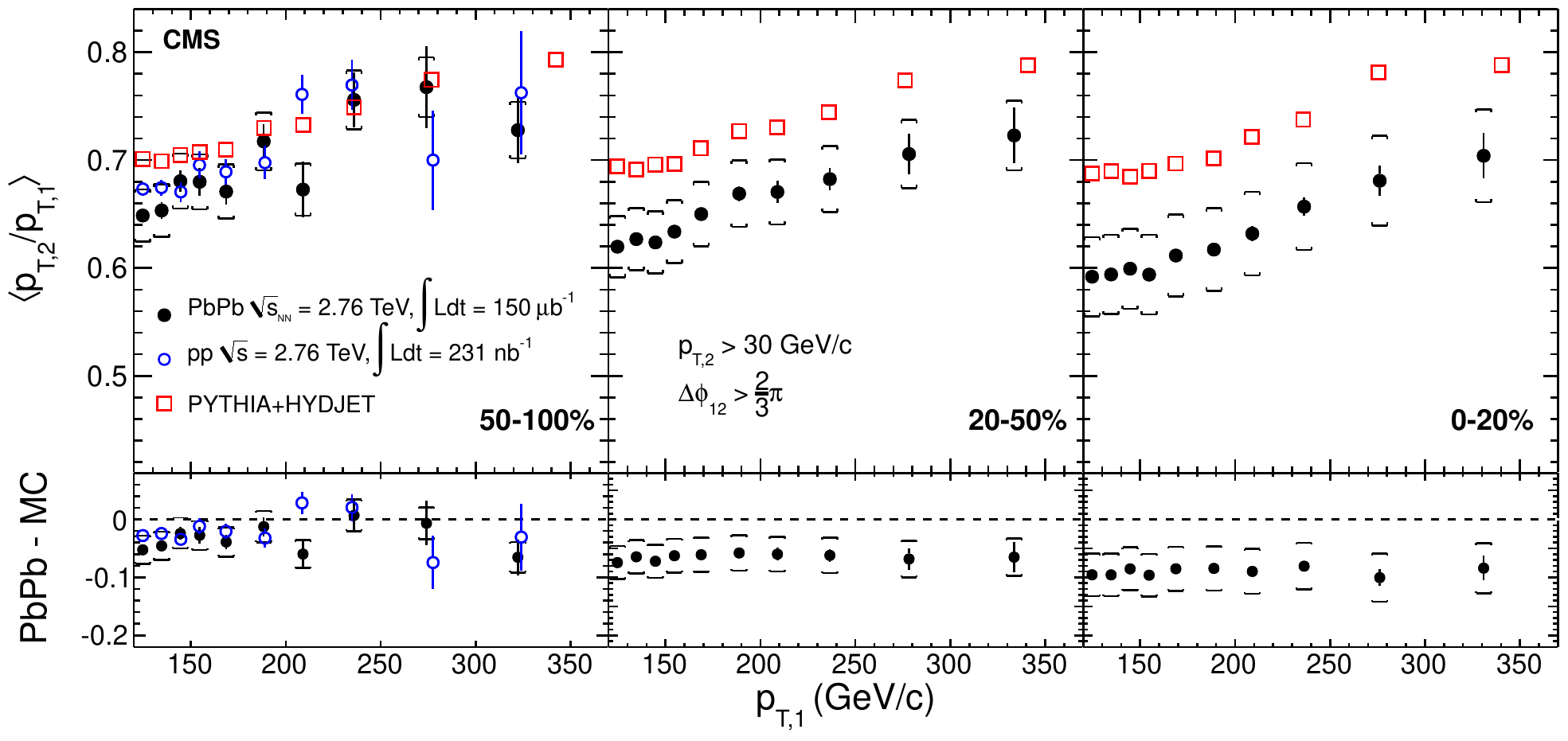}
\caption{
Average dijet momentum ratio $\ptsub/\ptlead$ as a function of
leading jet \pt for three bins of collision centrality, from peripheral to central collisions,
corresponding to selections of 50--100\%,  30--50\% and 0--20\%  of the total inelastic cross section.
Results for \PbPb\ data are shown as points with vertical bars and brackets indicating
the statistical and systematic uncertainties, respectively.  Results for \PYTHYD are shown as squares. In the 50--100\% centrality bin,
results are also compared with pp data, which is shown as the open circles.
The difference between the \PbPb\ measurement and the \PYTHYD expectations is shown in the bottom panels. }
\label{deltaPt}
\end{center}
\end{figure*}

As shown in Fig.~\ref{deltaPt}, both the PbPb data and the \PYTHYD  samples reveal an increasing trend for the mean value of the
 jet transverse momentum ratio, as a function of the leading jet $\ptlead$. This can be understood
by the reduction in the effects of jet splitting and energy resolution as one goes to higher jet momenta.
However, the central \PbPb\ data points lie consistently below the \PYTHYD trend. The difference between the pp data and the \PYTHYD reference is of the order of the systematic uncertainty of the measurement, whereas the difference between \PbPb\ data and the reference is more than twice larger. This difference is related to the parton energy loss and for central PbPb collisions it is of significant magnitude across the whole \pt\ range explored in this study.

\begin{table}[htbp]
\begin{center}
\caption{Summary of the \ptrat\ systematic uncertainties. The range of values represent the
variation from low ($\ptlead<140\GeVc$) to high
($\ptlead>300\GeVc$) leading jet \pt. }
\label{RatioSystematics} \begin{tabular}{|l|c|c|c|c|}
\hline
Source &  50--100\% & 20--50\% & 0--20\% \\
\hline
Underlying event & 1\% &  3\% & 5\% \\
Jet energy scale & 3\% &3\% & 3\%  \\
Jet efficiency  & 1--0.1\% & 1--0.1\% & 1--0.1\% \\
Jet misidentification & $<0.1$\%  & $<0.1$\% & 1--0.1\%\\
Calorimeter noise & $<0.1$\% & $<0.1$\% & $<0.1$\%  \\
Jet identification & $<0.1$\% & $<0.1$\% & $<0.1$\%\\
\hline
Total   & 3.5\% & 4.5\% & 6\% \\
\hline
\end{tabular}
\end{center}
\end{table}

\section{Summary}
\label{sec:summary}

Dijet production in PbPb collisions at
$\rootsNN=2.76\TeV$ was studied with the CMS detector in a data sample
corresponding to an integrated luminosity of \lum. The \ak\ algorithm was
used to reconstruct jets based on combined tracker and calorimeter
information. Events containing a leading jet with $\ptlead > 120$\GeVc\ and
a subleading jet with $\ptsub > 30$\GeVc\ in the pseudorapidity range
$|\eta| < 2$ were analyzed.
Data were compared to \PYTHYD dijet simulations,
tuned to reproduce the observed underlying event fluctuations.
For the most peripheral collisions, good agreement between data and simulations is
observed. For more central collisions, the dijet momentum imbalance in the data
is significantly larger than seen in the simulation. Across the entire range of jet momenta studied, no significant broadening of the dijet angular correlations is observed with respect to the reference distributions.

The dijet momentum imbalance was studied as a function of the leading jet $\ptlead$
for different centrality ranges in comparison to the \PYTHYD simulation.
For leading jet momenta
$\ptlead > 180$\GeVc\ the dijet balance distributions are found to be essentially
unbiased by the subleading jet threshold of $\ptsub> 30 $\GeVc.
For mid-central (30--50\%) and more central PbPb event selections, a significantly
lower average dijet momentum ratio $\langle \ptsub/\ptlead \rangle$
is observed than in the pp data and in the dijet embedded simulations. The downward shift in
$\langle \ptsub/\ptlead \rangle$, with respect to the \PYTHYD reference, is seen to increase
monotonically with  increasing collision centrality,
and to be largely independent of the leading jet $\ptlead$,
up to $\ptlead$ values in excess of 350\GeVc.

In summary, the results presented in this paper confirm previous observations based on
a smaller dataset and extend the measurements of jet-quenching effects to wider centrality and
leading jet transverse momentum ranges, as well as to lower subleading jet transverse momentum.

\section*{Acknowledgments}
\hyphenation{Bundes-ministerium Forschungs-gemeinschaft Forschungs-zentren} We congratulate our colleagues in the CERN accelerator departments for the excellent performance of the LHC machine. We thank the technical and administrative staff at CERN and other CMS institutes. This work was supported by the Austrian Federal Ministry of Science and Research; the Belgium Fonds de la Recherche Scientifique, and Fonds voor Wetenschappelijk Onderzoek; the Brazilian Funding Agencies (CNPq, CAPES, FAPERJ, and FAPESP); the Bulgarian Ministry of Education and Science; CERN; the Chinese Academy of Sciences, Ministry of Science and Technology, and National Natural Science Foundation of China; the Colombian Funding Agency (COLCIENCIAS); the Croatian Ministry of Science, Education and Sport; the Research Promotion Foundation, Cyprus; the Ministry of Education and Research, Recurrent financing contract SF0690030s09 and European Regional Development Fund, Estonia; the Academy of Finland, Finnish Ministry of Education and Culture, and Helsinki Institute of Physics; the Institut National de Physique Nucl\'eaire et de Physique des Particules~/~CNRS, and Commissariat \`a l'\'Energie Atomique et aux \'Energies Alternatives~/~CEA, France; the Bundesministerium f\"ur Bildung und Forschung, Deutsche Forschungsgemeinschaft, and Helmholtz-Gemeinschaft Deutscher Forschungszentren, Germany; the General Secretariat for Research and Technology, Greece; the National Scientific Research Foundation, and National Office for Research and Technology, Hungary; the Department of Atomic Energy and the Department of Science and Technology, India; the Institute for Studies in Theoretical Physics and Mathematics, Iran; the Science Foundation, Ireland; the Istituto Nazionale di Fisica Nucleare, Italy; the Korean Ministry of Education, Science and Technology and the World Class University program of NRF, Korea; the Lithuanian Academy of Sciences; the Mexican Funding Agencies (CINVESTAV, CONACYT, SEP, and UASLP-FAI); the Ministry of Science and Innovation, New Zealand; the Pakistan Atomic Energy Commission; the Ministry of Science and Higher Education and the National Science Centre, Poland; the Funda\c{c}\~ao para a Ci\^encia e a Tecnologia, Portugal; JINR (Armenia, Belarus, Georgia, Ukraine, Uzbekistan); the Ministry of Education and Science of the Russian Federation, the Federal Agency of Atomic Energy of the Russian Federation, Russian Academy of Sciences, and the Russian Foundation for Basic Research; the Ministry of Science and Technological Development of Serbia; the Ministerio de Ciencia e Innovaci\'on, and Programa Consolider-Ingenio 2010, Spain; the Swiss Funding Agencies (ETH Board, ETH Zurich, PSI, SNF, UniZH, Canton Zurich, and SER); the National Science Council, Taipei; the Scientific and Technical Research Council of Turkey, and Turkish Atomic Energy Authority; the Science and Technology Facilities Council, UK; the US Department of Energy, and the US National Science Foundation.

Individuals have received support from the Marie-Curie programme and the European Research Council (European Union); the Leventis Foundation; the A. P. Sloan Foundation; the Alexander von Humboldt Foundation; the Belgian Federal Science Policy Office; the Fonds pour la Formation \`a la Recherche dans l'Industrie et dans l'Agriculture (FRIA-Belgium); the Agentschap voor Innovatie door Wetenschap en Technologie (IWT-Belgium); the Council of Science and Industrial Research, India; and the HOMING PLUS programme of Foundation for Polish Science, cofinanced from European Union, Regional Development Fund.

\bibliography{auto_generated}   

\cleardoublepage \appendix\section{The CMS Collaboration \label{app:collab}}\begin{sloppypar}\hyphenpenalty=5000\widowpenalty=500\clubpenalty=5000\textbf{Yerevan Physics Institute,  Yerevan,  Armenia}\\*[0pt]
S.~Chatrchyan, V.~Khachatryan, A.M.~Sirunyan, A.~Tumasyan
\vskip\cmsinstskip
\textbf{Institut f\"{u}r Hochenergiephysik der OeAW,  Wien,  Austria}\\*[0pt]
W.~Adam, T.~Bergauer, M.~Dragicevic, J.~Er\"{o}, C.~Fabjan, M.~Friedl, R.~Fr\"{u}hwirth, V.M.~Ghete, J.~Hammer\cmsAuthorMark{1}, N.~H\"{o}rmann, J.~Hrubec, M.~Jeitler, W.~Kiesenhofer, M.~Krammer, D.~Liko, I.~Mikulec, M.~Pernicka$^{\textrm{\dag}}$, B.~Rahbaran, C.~Rohringer, H.~Rohringer, R.~Sch\"{o}fbeck, J.~Strauss, A.~Taurok, F.~Teischinger, P.~Wagner, W.~Waltenberger, G.~Walzel, E.~Widl, C.-E.~Wulz
% ...
\vskip\cmsinstskip
\dag:~Deceased\\
1:~~Also at CERN, European Organization for Nuclear Research, Geneva, Switzerland\\
% ...

\end{sloppypar}
\end{document}